# Computational Drug Repositioning and Elucidation of Mechanism of Action of Compounds against SARS-CoV-2

**Francesco Napolitano**[1], **Gennaro Gambardella**[2,3], **Xin Gao**[1], **and Diego di Bernardo**[2,3,*]

[1]Computational Bioscience Research Center, King Abdullah University of Science and Technology (KAUST), Thuwal 23955-6900, Saudi Arabia
[2]Telethon Institute of Genetics and Medicine (TIGEM), Pozzuoli (NA) 80078, Italy and
[3]Department of Chemical, Materials and Industrial Production Engineering, University of Naples Federico II, 80125 Naples, Italy
[*]dibernardo@tigem.it

## ABSTRACT

The COVID-19 crisis called for rapid reaction from all the fields of biomedical research. Traditional drug development involves time consuming pipelines that conflict with the urgency of identifying effective therapies during a health and economic emergency. Drug repositioning, that is the discovery of new clinical applications for drugs already approved for different therapeutic contexts, could provide an effective shortcut to bring COVID-19 treatments to the bedside in a timely manner. Moreover, computational approaches can help accelerate the process even further. Here we present the application of computational drug repositioning tools based on transcriptomics data to identify drugs that are potentially able to counteract SARS-CoV-2 infection, and also to provide insights on their mode of action. We believe that mucolytics and HDAC inhibitors warrant further investigation. In addition, we found that the DNA Mismatch repair pathway is strongly modulated by drugs with experimental in vitro activity against SARS-CoV-2 infection. Both full results and methods are publicly available.

## 1 Introduction

Drug repositioning or drug repurposing aims to find a new clinical application for a drug already in use but for a different purpose. Usually, in drug repurposing, a known and potentially therapeutic target is selected and an experimental search for existing compounds able to modulate the target activity is performed. Computational drug repurposing offers a complementary approach to prioritise compounds for experimental validation. Several different methods have been proposed in the literature[1–3] and some have already been applied to SARS-CoV-2[4]. These can be broadly classified into methods searching for small molecules able to bind an active pocket starting from three-dimensional structure of the target protein[5] and those searching for compounds able to modulate the expression of the target mRNA[6,7]. While most of the existing drug repositioning pipelines focus on the former approach, here we used the latter to identify FDA approved drugs able to modulate the expression of genes expressed in the airway epithelium and that are known to interact with SARS-CoV-2 proteins. Next, we attempted a completely agnostic approach to identify drugs reversing the gene expression profile induced by SARS-CoV-2 infection. Finally, we investigated potential effective mechanisms of action by identifying molecular pathways that are consistently dysregulated by a set of 24 drugs recently proposed as effective to reduce SARS-CoV-2 infection in treated Vero cells. By application of the proper computational tools, we were thus able to identify a set of drugs that should be further experimentally validated and that may have a potential beneficial role in COVID19 treatment, together with insights about mechanisms of action that could help to identify the most effective targets against the infection.

## 2 Material & Methods

All the used computational tools share the same database of 6,100 drug induced gene expression profiles included in the Connectivity Map 2.0 (CMap)[8]. CMap data were produced using Affymetrix Micrarrays, which we pre-processed to abtain the expression values for 12,012 genes. Moreover, expression profiles induced by the same drug across different replicates and experimental conditions were merged together into a single consensus profile. Details of the preprocessing are reported in[9].

The Gene2drug[6] and Drug Set Enrichment Analysis (DSEA)[9] are two bioinformatic tools for drug-target prioritization and drug mode of action elucidation, respectively (see next Sections). They both use a pathway-based version of the CMap. A Pathway-based Expression Profiles (PEP) is obtained from a Gene Expression Profiles (GEP) by iteratively applying Gene Set Enrichment Analysis (GSEA[10]) to the GEP for each pathway in a database such as the Gene Ontology or KEGG. In particular, both tools use all the pathway collections included in the MSigDB v6.1[10], as previously published[9].

The publicly available online implementation of Gene2drug (http://gene2drug.tigem.it) was used to identify drugs downregulating ACE2 and TMPRSS2. The tool automatically identifies all the gene sets involving the input genes across all of the pathway databases. Similarly, for the DSEA analysis the online tool was used (http://dsea.tigem.it).

Since the Gene2drug analysis applied to the SARS-CoV-2 human molecular interactors required customization, it was performed offline using the gep2pep Bioconductor package[11]. In particular, all the human interactors of each for each of the 27 SARS-CoV-2 proteins[12] were defined as a gene-set. The 27 obtained gene sets were then added to each of the pathway collections included in the MSigDB v7[10]. Finally, the CMap GEPs were converted to PEPs based on these newly created gene set collections. In order to prioritize drugs according to their PEPs, Gene2drug analysis was performed for the 27 SARS-CoV-2 related gene sets, using all the MSigDB gene sets as statistical background. Since the statistical background was different for each gene set collection, the average scores were computed to obtain the final prioritization. The dataset of PEPs is publicly available[13].

Drugs predicted to reverse the SARS-CoV-2 induced signature were identified using the publicly avaliable MANTRA tool (http://mantra.tigem.it). RNA-seq raw counts for the identification of differentially expressed genes (DEGs) in response to SARS-CoV-2 were downloaded from GEO database (accession number GSE147507). Raw counts were first normalized using the edgeR tool while DEGs identified by using the limma package with its voom method in the R statistical environment 3.6. The obtained GEP was added to the MANTRA network using the "reverse node" feature. In this way, the GEP is reversed (most up-regulated genes appear as the most down-regulated and vice-versa) and the closest nodes in the network correspond to drugs inducing the opposite GEP as compared to that of SARS-CoV-2 infection.

Two-dimensional structure images and annotations for the compounds reported in the tables were obtained from the PubChem database[14].

## 3 Results and Discussion

### 3.1 Identification of drugs reducing the expression ACE2 and TMPRSS2

It has been recently shown tha SARS-CoV-2 entry in host cells is mediated by the ACE2 receptor and requires priming by the TMPRSS2 protease[12]. We thus sought to identify drugs that could potentially reduce the expression of both genes, although the benefit of such an approach is being debated[15]. To this end, we applied a computational drug repositioning approach named Gene2Drug (http://gene2drug.tigem.it)[6]. This tool computes an Enrichment Score (ES) and the corresponding P-value for each of 1309 small molecules of the Connectivity Map dataset (Broad Institute), including FDA approved drugs, based on how much they tend to down-regulate the genes of interest (ACE2 and TMPRSS2), as well as other genes involved in the same pathways. Results from the analysis identified 10 drugs shown in Table 1 ($p < 0.05$). Most of these drugs are potentially relevant: **carbenoxolone** has shown antiviral activity against the Dengue virus[16]; **indomethacin** is a non-steroidal anti-inflammatory drug (NSAID) inhibiting Prostaglandin E2 synthase (PTGES2) with a demonstrated efficacy against SARS-CoV[17]. Interestingly, in a recent study, SARS-CoV-2 viral NSP7 was found to interact with PTGES2[12] and thus the authors suggested indomethacin as potentially useful in treating patients. Here, however, indomethacin was selected by Gene2Drug because it appears to lower expression of genes in the ACE2 pathway, which does not include PTGES2 itself. Therefore indomethacin could be a high priority molecule to be tested as it could have a double effect. Another NSAID is present among the drugs identified in Table 1, **nimesulide** which could have a similar effect as indomethacin. Gene2Drug also identified **nicarpidine**, an angiotensin inhibitor, an obvious candidate but found using a purely data-driven approach.

### 3.2 Identification of drugs reducing expression of SARS-CoV-2 interactors

Another potentially effective approach to the identification of therapeutic agents counteracting SARS-CoV-2 infection is to target the host-virus interaction. In order to investigate cellular proteins interacting with viral proteins, we analyzed a set of 332 protein interactions between 26 SARS-CoV-2 proteins (plus 1 mutant) and human proteins obtained by performed an affinity purification-mass spectrometry analysis[12]. Since each of these interactions could be key to hijack the host during the course of infection, we sought to prioritize existing drugs potentially interfering with them. In particular, we applied once more Gene2drug[6], this time exploiting its gene-set wise analysis capability. Specifically, for each viral protein, we generated a corresponding gene-set containing all the host protein interactors, thus obtaining 27 gene-sets. We then applied Gene2Drug to identify those compounds that are able to downregulate the expression of most of the genes across the 27 gene-sets at the same time, which appears a particularly effective strategy given the currently limited understanding of the specific infection mechanisms. The top 20 hits are reported in Table 2. Interestingly, one of the hits is **niclosamide**, an antihelminthic drug, that was found to have antiviral efficacy against SARS-CoV-2 in a recent screening using VERO cells[18]. Other notable candidates found by our analysis are: **fenoterol**, which was also identified as potentially effective in a recent study on computational drug repositioning[19]; **alexidine**, an antibiotic and a selective inhibitor of the mitochondrial phosphatase Ptpmt1, that was found to inhibit replication of the CytoMegaloVirus (CMV) infection[20]; **oxytetracycline**, another antibiotic that has shown antiviral activity against Dengue virus[21] and the ability to reduce HIV-1 RNA transcripts within extracellular vescicles[22]; **clofibrate**, a peroxisome-proliferator activated receptor-alpha (PPAR-) agonist previously used as a cholesterol-lowering agent, which was found to have antiviral activity against MDV, an alphaherpesvirus that infects chickens and causes a deadly lymphoma[23]; **flupentixol**, a neuroleptic agent that was found to inhibit hepatitis C virus entry[24] and coxsackievirus replication[25]; **gossypol**, a natural phenol derived from the cotton plant known to inhibit spermatogenesis, that was shown to have also antiviral activity[26]; **ms-275** and **trichostatin A**, two histone deacetylases inhibitors (HDACi), the latter found to inhibit expression of herpes simplex virus genes[27]; interestingly valproic acid, another HDACi, was proposed for repositioning against SARS-CoV-2, as the viral protein NSP5 was found to interact with HDAC2[12].

### 3.3 Identification of drugs reversing the transcriptional signature induced by SARS-CoV-2 infection

A data-driven approach to drug repositioning is to identify compounds that are able to revert a disease-related signature[3]. We developed the MANTRA tool exploiting the cMap dataset (Broad Institute) to generate and explore drug networks for drug repositioning[28,29]. In drug networks, nodes represent drug-induced transcriptional profiles, and edges represent similarities between them. We sought to use this approach to prioritize drugs reverting SARS-CoV-2 induced gene expression at the whole-genome level without assuming any prior knowledge of molecular mechanisms involved in the disease progression. Recently, gene expression profiles of primary human bronchial epithelial (NHBE) infected with SARS-CoV-2 have been made publicly available[30] (GSE147507 from the GEO database). We obtained the corresponding raw data and computed the differential expression of 15,330 genes between infected and non infected cells. Out of these, only 37 genes were significantly differentially expressed (adjusted $p < 0.1$), most of which were over-expressed in infected cells and included interleukin 6 (IL6) and 8 (IL8), chemokines and interferon alpha inducible proteins. The MANTRA approach is based on Gene Set Enrichment Analysis and therefore it requires in input a list of all the measured genes ranked according to their differential expression, including both significant and non-significant genes. We thus sorted all the 15,330 genes according to their differential expression, but with genes most down-regulated following infection *at the top of the list* and those most up-regulated *at the bottom*. We then queried the MANTRA drug network with this profile and looked for drugs inducing a similar transcriptional profile, which therefore may potentially induce a profile opposite to that of the infection. We thus found 11 significant drugs (transcriptional distance $< 0.8$) listed in Table 3. One of the 11 drugs is sirolimus, also known as rapamycin, an inhibitor of mechanistic Target Of Rapamycin Complex 1 (mTORC1), and used clinically as an immunosuppressive agent. This is an obvious candidate for this kind of drug repositioning, where the aim is to find a drug inducing an opposite repsonse to that of the virus, as most of the genes overexpressed during viral infection are immune-related. Of course, this does not mean that it is clinically relevant as suppressing the immune response could be detrimental to the cells. Nevertheless, Rapamycin was also suggested in the SARS-CoV-2 interactors' study based on the nucleocapsid interaction with the mTOR translational repressors LARP1[12] and on rapamycin reported *in vitro* activity against MERS[31]. Other interesting candidates found by MANTRA are: **ambroxol** (a metabolite of bromhexine), a mucolytic agent with reported antiviral activity against influenza-virus among others[32,33] and with inhibition activity against TMPRSS2[34]; **corticosterone**, a glucocorticoid; **idoxuridine**, a nucleoside analogue used to inhibit replication of DNA viruses; **naltrexone**, a mu-opiod receptor antagonist used for opioid and alcohol dependence, with reported antiinflammatory effect[35]; **nordihydroguaiaretic acid** an antioxidant compound found in the creosote bush (Larrea tridentata) with reported antiviral activity against West Nile and Zika viruses[36].

### 3.4 Mechanism of action of compounds with reported *in vitro* antiviral action against SARS-CoV-2 infection.

Understanding the biological mechanisms underlying a pathological condition at the molecular level can greatly help the identification of potentially effective drugs. One way of obtaining such knowledge is the analysis of existing small molecules that have been observed to elicit varying levels of therapeutic efficacy through drug screening. Although the mode of action of positive hits from large-scale drug screening can be difficult to identify, it is likely that such hidden mechanism is shared among most of the hits. In order to investigate and exploit this assumption, we previously developed the Drug Set Enrichment Analysis (DSEA[9], http://dsea.tigem.it). DSEA is able to highlight molecular pathways that are consistently and specifically dysregulated by most drugs in a set by analyzing drug-induced expression profiles in the CMap dataset.

In order to identify compounds that are able to counteract SARS-CoV-2 infection, Jeon et al[18] recently screened a collection of 35 small molecules previously observed as potentially effective against SARS-CoV infection, plus 15 drugs suggested by infectious diseases specialists. Among the 50 screened drugs, 24 showed reduction of the SARS-CoV-2 N protein levels in infected VERO cells[18]. Of the 24 drugs, 10 were present in the cMap dataset used by DSEA (amodiaquine, ciclosporin, desoxycortone, digoxin, loperamide, mefloquine, niclosamide, ouabain, proscillaridin, tetrandrine). We thus sought to analyze these 10 drugs through the DSEA tool in order to understand the common mechanism of action underlying their antiviral effect. These drugs were found by DSEA to modulate targets of the transcription factors (TFs) shown in Table 4, including: **NF-kB**, a master regulator of stress response; **ATF6**, a TF involved in the integrated stress response and with a known role in enhancing herpesvirus gene expression[37]; **STAT5B**, a member of the STAT family activated in response to cytokines and growth factor. DSEA also highlighted several biological processes as modulated by these drugs (Table 4), among which those related to DNA mistmatch repair pathway (MMR). Interestingly, DNA MMR pathway activation has been recently reported to be required for the replication of the coronavirus infection bronchitis virus (IBV)[38].

### 3.5 Conclusions

Computational methods for drug repositioning could narrow down the search for effective drugs to counteract novel epidemics. Here we showed how a number of existing tools can be used towards this aim, covering different approaches such as targeting of host proteins interacting with viral proteins, reversion of disease-induced transcriptional profiles, and mode of action investigation for drugs with demonstrated *in vitro* activity against SARC-CoV-2 infection. Overall, we found 39 compounds that could be tested experimentally among which NSAIDs, antihistamines, antibiotics and antihelmintics, antipsychotics, corticosteroids, mucolytics and HDAC inhibitors (HDACi). Bromhexine, of which ambroxol is a metabolite, has been recently proposed as a prophylactic treatment for SARS-COV-2 patients[39] based on its reported activity against TMPRSS2 and its over-the-counter availability and safety levels. In our study, we find that ambroxol also induces a transcriptional profile opposite to that induced by SARS-CoV-2 infection in vitro, adding to the evidence of a possible role of this drug in counteracting the virus. Of interest, HDAC2 was found to interact with SARS-CoV-2 proteins[12] and the screening on inhibitors of infection on VERO cells identified several members of the cardiac glycosides as effective[18]. Cardiac glycosides have been recently found to have strong HDACi activity[40,41]. It will be therefore interesting to verify experimentally whether HDACi have beneficial role in reducing viral replication in host cells. Finally, by interrogating the possible mode of action of positive hits of a drug screening in VERO cells[18], we identified the DNA MMR pathway as possibly involved in the mode of action of these drugs, and therefore hinting that this pathway could be involved in the host response to the viral infection. Interestingly, HDAC proteins are involved in the modulation of the DNA MMR pathway, thus hinting that HDAC inhibtors, if proven to be effective, may act through this pathway.

### 3.6 Data Availability

All the cited methods, the full lists of prioritized drugs and pathways, and all the necessary data to reproduce the results, are publicly available[13].

## 4 Funding

The research reported in this publication was supported by the King Abdullah University of Science and Technology (KAUST) Office of Sponsored Research (OSR) under Award No. FCC/1/1976-04, FCC/1/1976-06, FCC/1/1976-17, FCC/1/1976-18, FCC/1/1976-23, FCC/1/1976-25, FCC/1/1976-26, URF/1/3450-01, URF/1/4098-01-01, and REI/1/0018-01-01 to XG and by Fondazione Telethon grant to DdB.

## 5 Acknowledgements

We thank Diego Carrella of the TIGEM Bioinformatics Core for support with the MANTRA webtool.

## References

1. Iorio, F., Saez-Rodriguez, J. & di Bernardo, D. Network based elucidation of drug response: from modulators to targets. *BMC Syst. Biol.* **7**, 139, DOI: 10.1186/1752-0509-7-139 (2013).

2. Sirci, F., Napolitano, F. & di Bernardo, D. Computational Drug Networks: a computational approach to elucidate drug mode of action and to facilitate drug repositioning for neurodegenerative diseases. *Drug Discov. Today: Dis. Model.* **19**, 11–17, DOI: 10.1016/j.ddmod.2017.04.004 (2016).

3. Mottini, C., Napolitano, F., Li, Z., Gao, X. & Cardone, L. Computer-aided drug repurposing for cancer therapy: Approaches and opportunities to challenge anticancer targets. *Semin. Cancer Biol.* DOI: 10.1016/j.semcancer.2019.09.023 (2019).

4. Zhou, Y. *et al.* Network-based drug repurposing for novel coronavirus 2019-nCoV/SARS-CoV-2. *Cell Discov.* **6**, DOI: 10.1038/s41421-020-0153-3 (2020).

5. Lavecchia, A. & Di Giovanni, C. Virtual Screening Strategies in Drug Discovery: A Critical Review. .

6. gene2drug: a computational tool for pathway-based rational drug repositioning. *Bioinformatics* **34**, 1498–1505, DOI: 10.1093/bioinformatics/btx800 (2018).

7. Keenan, A. B. *et al.* The Library of Integrated Network-Based Cellular Signatures NIH Program: System-Level Cataloging of Human Cells Response to Perturbations. *Cell Syst.* **6**, 13–24, DOI: 10.1016/J.CELS.2017.11.001 (2018).

8. Lamb, J. *et al.* The Connectivity Map: using gene-expression signatures to connect small molecules, genes, and disease. *Sci. (New York, N.Y.)* **313**, 1929–35, DOI: 10.1126/science.1132939 (2006).

9. Napolitano, F., Sirci, F., Carrella, D. & Di Bernardo, D. Drug-Set Enrichment Analysis: A Novel Tool to Investigate Drug Mode of Action. *Bioinformatics* **32**, 235–241, DOI: 10.1093/bioinformatics/btv536 (2016).

10. Subramanian, A. *et al.* Gene set enrichment analysis: A knowledge-based approach for interpreting genome-wide expression profiles. *Proc. Natl. Acad. Sci.* **102**, 15545–15550 (2005).

11. Napolitano, F., Carrella, D., Gao, X. & Bernardo, D. gep2pep: a Bioconductor package for the creation and analysis of pathway-based expression profiles. *Bioinformatics* (2019).

12. Gordon, D. E. *et al.* A SARS-CoV-2-Human Protein-Protein Interaction Map Reveals Drug Targets and Potential Drug-Repurposing. *bioRxiv* DOI: 10.1101/2020.03.22.002386 (2020).

13. Napolitano, F., Gambardella, G., Carrella, D., Gao, X. & di Bernardo, D. Computational Drug Repositioning and Elucidation of Mechanism of Action of Compounds against SARS-CoV-2 - Supplementary data, DOI: 10.5281/zenodo.3756887 (2020).

14. Kim, S. *et al.* PubChem 2019 update: Improved access to chemical data. *Nucleic Acids Res.* **47**, D1102–D1109, DOI: 10.1093/nar/gky1033 (2019).

15. Mcmurray, J. J. V., Pfeffer, M. A., Ph, D. & Solomon, S. D. Renin – Angiotensin – Aldosterone System Inhibitors in Patients with Covid-19. *The New Engl. J. Medicine* 1–7 (2020).

16. Pu, J. *et al.* Antiviral activity of Carbenoxolone disodium against dengue virus infection. *J. Med. Virol.* **89**, 571–581, DOI: 10.1002/jmv.24571 (2017).

17. Amici, C. *et al.* Indomethacin has a potent antiviral activity against SARS coronavirus. *Antivir. Ther.* **11**, 1021–1030 (2006).

18. Jeon, S. *et al.* Identification of antiviral drug candidates against SARS-CoV-2 from FDA-approved drugs. *bioRxiv* DOI: 10.1101/2020.03.20.999730 (2020).

19. Wu, C. *et al.* Analysis of therapeutic targets for SARS-CoV-2 and discovery of potential drugs by computational methods. *Acta Pharm. Sinica B* DOI: 10.1016/j.apsb.2020.02.008 (2020).

20. Drug Repurposing Approach Identifies Inhibitors of the Prototypic Viral Transcription Factor IE2 that Block Human Cytomegalovirus Replication. *Cell Chem. Biol.* **23**, 340–351, DOI: 10.1016/j.chembiol.2015.12.012 (2016).

21. Rothan, H. A. *et al.* Study the antiviral activity of some derivatives of tetracycline and non-steroid anti inflammatory drugs towards dengue virus. *Trop. biomedicine* **30**, 681–90 (2013).

22. Antiretroviral Drugs Alter the Content of Extracellular Vesicles from HIV-1-Infected Cells. *Sci. Reports* **8**, 1–20, DOI: 10.1038/s41598-018-25943-2 (2018).

23. Boodhoo, N., Kamble, N., Sharif, S. & Behboudi, S. Glutaminolysis and Glycolysis Are Essential for Optimal Replication of Marek's Disease Virus. *J. Virol.* **94**, 1–13, DOI: 10.1128/jvi.01680-19 (2019).

24. Pietschmann, T. Clinically Approved Ion Channel Inhibitors Close Gates for Hepatitis C Virus and Open Doors for Drug Repurposing in Infectious Viral Diseases. *J. Virol.* **91**, 1–9, DOI: 10.1128/jvi.01914-16 (2017).

25. Zuo, J. *et al.* Fluoxetine is a potent inhibitor of coxsackievirus replication. *Antimicrob. Agents Chemother.* **56**, 4838–4844, DOI: 10.1128/AAC.00983-12 (2012).

26. Dodou, K. Investigations on gossypol: Past and present developments, DOI: 10.1517/13543784.14.11.1419 (2005).

27. Shapira, L., Ralph, M., Tomer, E., Cohen, S. & Kobiler, O. Histone deacetylase inhibitors reduce the number of herpes simplex virus-1 genomes initiating expression in individual cells. *Front. Microbiol.* **7**, 1970–1970, DOI: 10.3389/fmicb.2016.01970 (2016).

28. Discovery of drug mode of action and drug repositioning from transcriptional responses. *Proc. Natl. Acad. Sci.* **107**, 14621–14626, DOI: 10.1073/pnas.1000138107 (2010).

29. Mantra 2.0: An online collaborative resource for drug mode of action and repurposing by network analysis. *Bioinformatics* DOI: 10.1093/bioinformatics/btu058 (2014).

30. Blanco-Melo, D. *et al.* SARS-CoV-2 launches a unique transcriptional signature from in vitro, ex vivo, and in vivo systems. *bioRxiv* DOI: 10.1101/2020.03.24.004655 (2020).

31. Antiviral potential of ERK/MAPK and PI3K/AKT/mTOR signaling modulation for Middle East respiratory syndrome coronavirus infection as identified by temporal kinome analysis. *Antimicrob. Agents Chemother.* **59**, 1088–1099, DOI: 10.1128/AAC.03659-14 (2015).

32. Yang, B. *et al.* Ambroxol suppresses influenza-virus proliferation in the mouse airway by increasing antiviral factor levels. *Eur. Respir. J.* **19**, 952–958, DOI: 10.1183/09031936.02.00253302 (2002).

33. Paleari, D., Rossi, G. A., Nicolini, G. & Olivieri, D. Ambroxol: A multifaceted molecule with additional therapeutic potentials in respiratory disorders of childhood, DOI: 10.1517/17460441.2011.629646 (2011).

34. Shen, L. W., Mao, H. J., Wu, Y. L., Tanaka, Y. & Zhang, W. TMPRSS2: A potential target for treatment of influenza virus and coronavirus infections, DOI: 10.1016/j.biochi.2017.07.016 (2017).

35. Patten, D. K., Schultz, B. G. & Berlau, D. J. The Safety and Efficacy of Low-Dose Naltrexone in the Management of Chronic Pain and Inflammation in Multiple Sclerosis, Fibromyalgia, Crohn's Disease, and Other Chronic Pain Disorders. *Pharmacother. The J. Hum. Pharmacol. Drug Ther.* **38**, 382–389, DOI: 10.1002/phar.2086 (2018).

36. Merino-Ramos, T., Jiménez De Oya, N., Saiz, J. C. & Martín-Acebes, M. A. Antiviral activity of nordihydroguaiaretic acid and its derivative tetra-O-methyl nordihydroguaiaretic acid against West Nile virus and Zika virus. *Antimicrob. Agents Chemother.* **61**, DOI: 10.1128/AAC.00376-17 (2017).

37. Hinte, F., van Anken, E., Tirosh, B. & Brune, W. Repression of viral gene expression and replication by the unfolded protein response effector XBP1u. *eLife* **9**, DOI: 10.7554/eLife.51804 (2020).

38. Coronavirus infection induces DNA replication stress partly through interaction of its nonstructural protein 13 with the p125 subunit of DNA polymerase $\delta$. *J. Biol. Chem.* **286**, 39546–39559, DOI: 10.1074/jbc.M111.242206 (2011).

39. Depfenhart, M. *et al.* A SARS-CoV-2 Prophylactic and Treatment ; A Counter Argument Against The Sole Use of Chloroquine. *Am. J. Biomed. Sci. & Res.* **8**, 248–251, DOI: 10.34297/AJBSR.2020.08.001283 (2020).

40. Raynal, N. J. *et al.* Targeting calcium signaling induces epigenetic reactivation of tumor suppressor genes in cancer. *Cancer Res.* **76**, 1494–1505, DOI: 10.1158/0008-5472.CAN-14-2391 (2016).

41. Sirci, F. *et al.* Comparing structural and transcriptional drug networks reveals signatures of drug activity and toxicity in transcriptional responses. *npj Syst. Biol. Appl.* **3**, 1–12, DOI: 10.1038/s41540-017-0022-3 (2017).

| # | Drug | ES | PV | 2D structure | Notes |
|---|------|-----|------|--------------|-------|
| 1 | moracizine | -1.000 | 0.0092 | 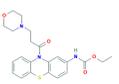 | Has a role as an anti-arrhythmia drug. |
| 2 | Gly-His-Lys | -0.995 | 0.0184 | 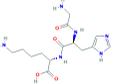 | Has a role as a metabolite, a chelator, a vulnerary and a hepatoprotective agent. |
| 3 | carbenoxolone | -0.995 | 0.0184 | 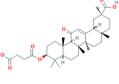 | Used for the treatment of digestive tract ulcers. |
| 4 | indometacin | -0.995 | 0.0184 | 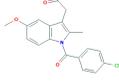 | Non-steroidal anti-inflammatory drug most commonly used in rheumatoid arthritis, ankylosing spondylitis, osteoarthritis, acute shoulder pains, and acute gouty arthritis. |
| 5 | vitexin | -0.995 | 0.0184 | 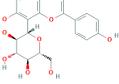 | platelet aggregation inhibitor, an alpha-glucosidase inhibitor, an antineoplastic agent and a plant metabolite |
| 6 | nimesulide | -0.991 | 0.0276 | 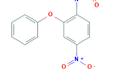 | Nonsteroidal anti-inflammatory drug, modestly selective COX-2 inhibitor. |
| 7 | PNU-0293363 | -0.981 | 0.0461 | N/A | N/A |
| 8 | chloropyramine | -0.981 | 0.0461 | 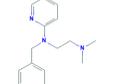 | Antihistamine drug with applications to the treatment of allergic conjunctivitis, allergic rhinitis, bronchial asthma, and other atopic (allergic) conditions. |
| 9 | nicardipine | -0.981 | 0.0461 | 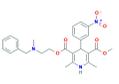 | Calcium channel blocker used in the treatment of hypertension and stable angina pectoris. |
| 10 | remoxipride | -0.981 | 0.0461 | 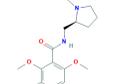 | Antipsychotic agent specific for dopamine D2 receptors, effective in the treatment of schizophrenia. |

**Table 1.** Drugs predicted to downregulate ACE2 and TMPRSS2 genes by the Gene2drug tool ($p < 0.05$). Enrichment score (ES) and the corresponding p-value (PV) are reported for each drug.

| #  | Drug             | mean ES | mean PV | 2D structure | Notes |
|----|------------------|---------|---------|--------------|-------|
| 1  | proxymetacaine   | -0.536  | 0.0002  | 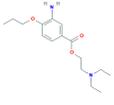 | Benzoic acid derivative anesthetic agent, with local anesthetic activity. |
| 2  | fenoterol        | -0.492  | 0.0008  | 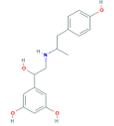 | Beta2-adrenergic agonist used in the management of reversible airway obstruction. |
| 3  | 0179445_0000     | -0.498  | 0.0013  | N/A          | N/A |
| 4  | cefepime         | 0.473   | 0.0015  | 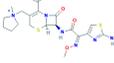 | Antibacterial drug. |
| 5  | picrotoxinin     | -0.473  | 0.0023  | 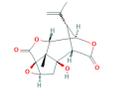 | Has a role as a plant metabolite, a GABA antagonist and a serotonergic antagonist. |
| 6  | alexidine        | -0.485  | 0.0036  | 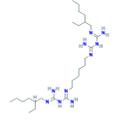 | Amphipathic bisbiguanide with a role as an antibacterial agent. |
| 7  | oxytetracycline  | -0.460  | 0.0042  | 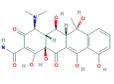 | Tetracycline used for treatment of infections caused by a variety of Gram positive and Gram negative microorganisms. |
| 8  | clofibrate       | -0.446  | 0.0052  | 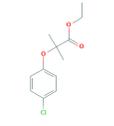 | Anticholesteremic drug antilipemic drug. |
| 9  | ms 275           | -0.439  | 0.0052  | 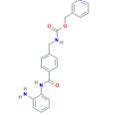 | Inhibitor of histone deacetylase isoform 1 (HDAC1) and isoform 3 (HDAC3). |
| 10 | flupentixol      | -0.356  | 0.0073  | 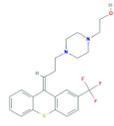 | Antipsychotic neuroleptic drug, used in schizophrenia. |
| 11 | n6 methyladenosine | -0.421 | 0.0084 | 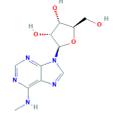 | methylated adenine residue |
| 12 | gossypol         | -0.398  | 0.0086  | 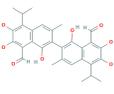 | Orally-active polyphenolic aldehyde with potential antineoplastic activity by inhibiting DNA replication and inducing apoptosis. |

| | | | | |
|---|---|---|---|---|
| 13 | dihydroergocristine | -0.41 | 0.0087 | 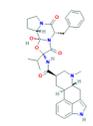 | Used as the mesylate salt for the symptomatic treatment of mental deterioration associated with cerebrovascular insufficiency and in peripheral vascular disease. It has a role as an adrenergic antagonist and a vasodilator agent. |
| 14 | suloctidil | -0.398 | 0.0089 | 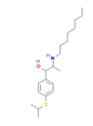 | An alkylbenzene. |
| 15 | trichostatin A | -0.389 | 0.0093 | 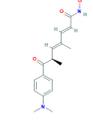 | Inhibits histone deacetylases. Potent inducer of tumor cell growth arrest, differentiation and apoptosis. |
| 16 | ticlopidine | -0.447 | 0.0093 | 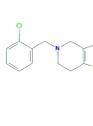 | Has a role as a fibrin modulating drug, a hematologic agent, an anticoagulant, a platelet aggregation inhibitor and a P2Y12 receptor antagonist. |
| 17 | cloperastine | -0.461 | 0.0098 | 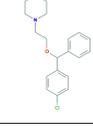 | A cough suppressant that acts on the central nervous system. |
| 18 | niclosamide | -0.401 | 0.0103 | 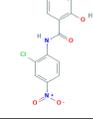 | A member of benzamides with anthelmintic and potential antineoplastic activity. |
| 19 | methylbenzethonium chloride | -0.454 | 0.0104 | 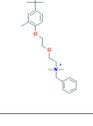 | N/A |
| 20 | clomipramine | -0.397 | 0.0106 | 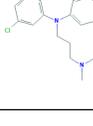 | Has a role as a serotonergic antagonist, a serotonergic drug, a serotonin uptake and trypanothione-disulfide reductase inhibitor, and an antidepressant. |

**Table 2.** Drugs predicted to downregulate SARS-CoV-2 human interactors by the Gene2drug tool ($p < 0.05$, top 20 reported). The mean enrichment score (ES) and the corresponding mean p-value (PV) across different gene-set backgrouns (see Methods) are reported for each drug.

| # | Drug | 2D structure | Notes |
|---|---|---|---|
| 1 | ambroxol | 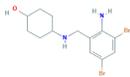 | Aromatic amine used in the treatment of respiratory diseases associated with viscid or excessive mucus. |
| 2 | amprolium | 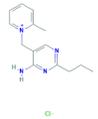 | A veterinary coccidiostat that interferes with thiamine metabolism. |
| 3 | benzamil | 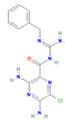 | Potent blocker of the ENaC channel and also a sodium-calcium exchange blocker. |
| 4 | chlorzoxazone | 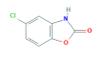 | Centrally acting muscle relaxant with sedative properties used for the symptomatic treatment of painful muscle spasm. |
| 5 | corticosterone | 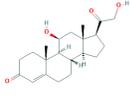 | Steroid hormone of the corticosteroid type. |
| 6 | doxylamine | 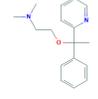 | A first generation ethanolamine with antiinflammatory, sedative and antihistamine properties. |
| 7 | idoxuridine | 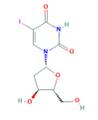 | Antiviral drug and DNA synthesis inhibitor, with antiviral activity against herpes simplex virus (HSV) and potential radiosensitizing activities. |
| 8 | meptazinol | 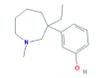 | Opioid analgesic. |
| 9 | naltrexone | 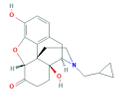 | A mu-opioid receptor antagonist used to treat alcohol dependence. Has also a role as a central nervous system depressant, an environmental contaminant, a xenobiotic and an antidote to opioid poisoning. |
| 10 | nordihydroguaiaretic acid | 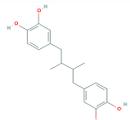 | Has a role as an antioxidant and a plant metabolite. |
| 11 | sirolimus | 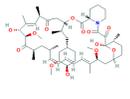 | Antibiotic, has a role as immunosupressive, antineoplastic, antibacterial agent, mTOR inhibitor and a bacterial metabolite. |

**Table 3.** Drugs predicted to revert the transcriptional signature induced by SARS-CoV-2 infection through the MANTRA tool (distance to the inverse of the SARS-CoV-2 profile $< 0.8$.

| Transcription Factor / GO Term | ES | PV |
|---|---|---|
| Transcription Factor Targets | | |
| NF-$\kappa$B (V$NFKAPPAB65_01) | 0.704 | 1.08E-04 |
| ATF6 (V$ATF6_01) | 0.675 | 2.35E-04 |
| OCT (V$OCT_Q6) | 0.669 | 2.77E-04 |
| SREBP1 (TCANNTGAY_V$SREBP1_01) | 0.668 | 2.82E-04 |
| NF-$\kappa$B (V$NFKAPPAB_01) | 0.667 | 2.94E-04 |
| OCT (V$OCT_C) | 0.665 | 3.07E-04 |
| STAT5B (V$STAT5B_01) | 0.658 | 3.68E-04 |
| NF-$\kappa$B (V$NFKB_Q6) | 0.656 | 3.89E-04 |
| SREBP1 (V$SREBP1_01) | 0.648 | 4.75E-04 |
| PITX2 (V$PITX2_Q2) | 0.646 | 5.05E-04 |
| Gene Ontology - Biological Process | | |
| RNA modification | -0.731 | 5.01E-05 |
| response to ischemia | -0.707 | 9.91E-05 |
| DNA replication | -0.682 | 1.95E-04 |
| amino acid transport | 0.681 | 2.03E-04 |
| DNA strand elongation involved in DNA replication | -0.679 | 2.12E-04 |
| intra-Golgi vesicle-mediated transport | 0.678 | 2.16E-04 |
| porphyrin-containing compound biosynthetic process | -0.678 | 2.16E-04 |
| chlorophyll biosynthetic process | -0.674 | 2.46E-04 |
| photosynthesis | -0.674 | 2.46E-04 |
| negative regulation of neuron projection development | -0.660 | 3.54E-04 |
| Gene Ontology - Molecular Function | | |
| MutLalpha complex binding | -0.795 | 7.08E-06 |
| polypeptide N-acetylgalactosaminyltransferase activity | -0.726 | 5.72E-05 |
| mismatched DNA binding | -0.701 | 1.17E-04 |
| nucleotide binding | -0.686 | 1.76E-04 |
| magnesium chelatase activity | -0.674 | 2.46E-04 |
| single-stranded DNA binding | -0.668 | 2.82E-04 |
| receptor activity | 0.659 | 3.61E-04 |
| antigen binding | 0.630 | 7.64E-04 |
| enzyme regulator activity | -0.629 | 7.76E-04 |
| ribonuclease P activity | -0.611 | 1.22E-03 |
| Gene Ontology - Cellular Component | | |
| nucleolus | -0.715 | 7.94E-05 |
| DNA replication factor C complex | -0.675 | 2.35E-04 |
| Gemini of coiled bodies | -0.654 | 4.08E-04 |
| Cul3-RING ubiquitin ligase complex | -0.638 | 6.27E-04 |
| nucleosome | 0.608 | 1.29E-03 |
| apical plasma membrane | 0.601 | 1.52E-03 |
| mitochondrial small ribosomal subunit | -0.596 | 1.73E-03 |
| extracellular matrix | 0.596 | 1.74E-03 |
| nucleolar ribonuclease P complex | -0.584 | 2.31E-03 |
| COPI vesicle coat | 0.577 | 2.72E-03 |

**Table 4.** Drug Set Enrichment Analysis results for 24 small molecules showed to lower the level of the N protein in SARS-CoV-2. The reported pathways are consistently dysregulated by most drugs in the set as found by the DSEA based on the analysis of their induced transcriptomes. Top 10 pathways are reported for the Transcription Factor Targets collection and for the three Gene Ontology collections. Enrichment Scores (ES) and the corresponding p-value (PV) are reported for each gene set.